\documentclass[12pt]{article}
\usepackage{graphics}
\usepackage{epsfig}
\usepackage[cp1251]{inputenc}
\usepackage{amssymb,latexsym,amsmath}
\newcommand{\beq}{\begin{equation}}
\newcommand{\eeq}{\end{equation}}
\newcommand{\bea}{\begin{eqnarray}}
\newcommand{\eea}{\end{eqnarray}}
\newcommand{\eps}{\varepsilon}

\newcommand{\Fs}{\mbox{\scriptsize F}}
\newcommand{\pr}{_{\perp}}
\newcommand{\bk}{{\bf k}}

\newcommand{\br}{{\bf r}}
\newcommand{\Vef}{V^{p}_{\mbox{\scriptsize eff}}}
\def\LPA{V^{\mbox{\scriptsize LPA}}_{\mbox{\scriptsize eff}}}
\newcommand{\DF}{\Delta_{\mbox{\scriptsize F}}}
\newcommand{\VF}{{\mathcal V}^{\mbox{\scriptsize F}}_{\mbox{\scriptsize eff}}}
\newcommand{\Vinf}{ V_{\rm eff}^{\rm inf}}

\begin{document}

\begin{center}
{\Large\bf The microscopic pairing gap in a slab of nuclear matter
for the Argonne v$_{18}$ $NN$-potential}
\end{center}
\vskip 1 cm

\centerline{S.~S.~Pankratov$^{a}$, M.~Baldo$^{b}$, U.~Lombardo$^{c,d}$,
E.E.~Saperstein$^{a}$ and M~.V.~Zverev$^{a}$}

\vskip 0.5 cm
\begin{small}
\centerline {$^{a}$ Kurchatov Institute, 123182, Moscow,
Russia }
\centerline{$^b$INFN, Sezione di Catania, 64 Via S.-Sofia, I-95123
Catania, Italy}
 \centerline{$^c$INFN-LNS, 44 Via S.-Sofia, I-95123
Catania, Italy}
\centerline{$^d$ 44 Via S.-Sofia, I-95123 Catania,
Italy}
\end{small}


\vskip 0.5 cm

\begin{abstract}
{\it Ab initio} gap equation for $^1S_0$ pairing in a nuclear slab
is solved for the Argonne v$_{18}$ $NN$-potential. The gap
function is compared in detail with the one found previously for
the separable form of the Paris potential. The difference between
the two gaps turned out to be about 10\%. Dependence of the gap on
the chemical potential $\mu$ is analyzed.

\end{abstract}
\newpage
\section{Introduction}

In the last decade, a progress was reached in the microscopic
theory of pairing in atomic nuclei. In this connection, the
studies of the Milan group \cite{milan1,milan2} should be cited
first. In the recent paper \cite{milan2} the direct solution of
the {\it ab initio} gap equation for a nucleus $^{120}$Sn was
carried out using the realistic Argonne v$_{14}$ $NN$-potential. A
bit later we made a similar calculation  \cite{Pankr} for a slab
of nuclear matter with parameters mimicking the atomic nuclei of
the tin region with the Paris potential \cite{Paris}. The
comparison of results of these two calculations rises some
questions. In Ref.~\cite{milan2} the gap $\Delta$, obtained by
solving the gap equation, turned out to be about a half of the
experimental value. The discrepancy was attributed  to the missing
contribution of the low-lying surface vibrations\cite{milan2}. On
the other hand, the gap found in \cite{Pankr} is much closer to
the experimental one and leaves a room to the surface vibration
corrections only for a 20\% about.

To some extent, the difference could originate from the difference
between the two systems under consideration, the spherical nucleus
and the slab. This reason seems to us  not very important as far
as only the curvature effects make the difference. Different
values of the effective nucleon mass $m^*$ used in these two
alternative calculations look  more essential. Indeed, it is well
known \cite{Bal} that the gap value in infinite matter is rather
sensitive to the value of the effective mass $m^*$. In
\cite{milan2} the coordinate dependent effective mass $m^*(r)$ was
used of the Sly4 version \cite{Sly4} of the Skyrme force. Its
value changes from $m^*=0.7m$ inside a nucleus to $m^*=m$ outside.
In \cite{Pankr} the bare nucleon mass, $m^*=m$, was used, in
accordance with prescriptions of \cite{KhS} and \cite{Fay}. The
specific form of the realistic $NN$-potential used in the gap
equation is one more possible reason of the difference under
discussion. In this article we analyze the latter point repeating
the calculation of \cite{Pankr}, but for the Argonne v$_{18}$
potential \cite{A18} which is quite similar to that of
\cite{milan2} but essentially different from the Paris potential.
Indeed, both the version of  the Argonne force are rather soft
core whereas the Paris potential is one of the most hard core
realistic $NN$-potentials. In addition, we examine the dependence
of the pairing gap on the chemical potential $\mu$ of the system
under consideration.

\section{The formalism}

The general method for solving the gap equation in a system with
the 1D inhomogeneity is developed in Ref.~\cite{Pankr}. Here we
present only the formalism which is necessary to describe the
procedure. Just as in Ref~\cite{Pankr}, we use the two-step method
\cite{Rep} to solve the microscopic gap equation,
\beq\Delta = {\cal V} A^{s}\Delta,\label{del}\eeq
where ${\cal V}$ is the free  $NN$-potential and $A^s=GG^s$ stands
for the two-particle propagator in the superfluid system. Here $G$
and $G^s$ are the one-particle Green functions without and with
pairing effects taken into account, respectively. In
Eq.~(\ref{del}), as usual, integration over intermediate
coordinates and summation over spin variables is understood. In
the nuclear matter theory, the gap equation in the form of
Eq.~(\ref{del})  is usually referred to as the
Bardeen--Cooper--Schrieffer (BCS) approximation.\footnote {In
physics of finite nuclei this term is commonly used for an
approximate way to solve the pairing problem, contrary to solving
the Bogolyubov equations.} In the more profound many-body theory
\cite{AB,Schuck} the irreducible block of the $NN$-interaction
should replace ${\cal V}$ in Eq.~(\ref{del}).

The complete Hilbert space $S$ of two-particle states is split
into two parts, $S=S_0+S'$. The first one is the model subspace
$S_0$ in which the gap equation is considered, and the other is
the complementary subspace $S'$. They are separated by the energy
$E_0$ in such a way that $S_0$ involves all the two-particle
states $(\lambda ,\lambda')$ with the single-particle energies
$\eps_{\lambda}, \eps_{\lambda'} <E_0$. The complementary subspace
$S'$ involves such two-particle states for which one of the
energies $\eps_{\lambda},\eps_{\lambda'}$ or both of them are
greater than $E_0$. Therefore, pairing effects can be neglected in
$S'$ if $E_0$ is sufficiently large. Validity of the inequality
$\Delta^2/(E_0-\mu)^2\ll 1$ is the criterium of such
approximation.

Correspondingly, the two-particle propagator is represented as the
sum $A^s = A^s_0 + A'$. Here we already neglected  the superfluid
effects in the $S'$-subspace and omitted the superscript ``s'' in
the second term. The gap equation (\ref{del}) can be rewritten in
the model subspace,
\beq \Delta = \Vef\,A^{s}_{0}\,\Delta\,,\label{del0}\eeq where the
effective pairing interaction (EPI) should be found in the
supplementary subspace, \beq\Vef = {\cal V} + {\cal V}
A'\,\Vef\,.\label{EPI}\eeq

As is well known, the gap equation (\ref{del}) can be written in
an equivalent form,
\beq \Delta = - {\cal V} \varkappa, \label{delkap} \eeq where the
anomalous density matrix $\varkappa$ is expressed directly in
terms of the Bogolyubov $u$-, $v$-functions:
\beq \varkappa({\bf r}_1,{\bf r}_2) = \sum_i u_i({\bf r}_1)
v_i({\bf r}_2). \label{kapuv} \eeq The summation in (\ref{kapuv})
is carried out over the complete set of the Bogolyubov functions
with eigenenergies $E_i>0$.

The gap equation (\ref{del0}) in the model subspace can be also
written in a form similar to (\ref{delkap}),
\beq \Delta = -{\Vef}\varkappa_0\,,\label{gap_kap}\eeq
where $\varkappa_0$ is the anomalous density matrix specified in
the model subspace as
\beq\varkappa_0(\br_1,\br_2) =  \int
A^s_0(E;\br_1,\br_2,\br_3,\br_4)\Delta(\br_3,\br_4) d\br_3 d\br_4
\,.\label{kap0}\eeq

However, a simple expression for this quantity similar to
Eq.~(\ref{kapuv}) does not exist. To obtain the explicit form of
$\varkappa_0$ it is convenient to use the basis of the
single-particle functions $\phi_{\lambda}({\bf r})$ which makes
the normal Green function $G$ diagonal:
 $G_{\lambda\lambda'}(\eps){=}G_{\lambda}(\eps) \delta_{\lambda\lambda'}$.
 Then one gets:
\beq u_i({\bf r}) = \sum_{\lambda} u_i^{\lambda}
\phi_{\lambda}({\bf r}), \label{ulam} \eeq
\beq v_i({\bf r}) = \sum_{\lambda} v_i^{\lambda}
\phi_{\lambda}({\bf r}), \label{vlam} \eeq
\beq \Delta ({\bf r}_1,{\bf r}_2) = \sum_{\lambda,\lambda'}
\Delta_{\lambda\lambda'} \phi_{\lambda}({\bf r}_1)
\phi_{\lambda'}({\bf r}_2). \label{delam} \eeq

After a simple algebra \cite{Pankr}, one finds
 \beq \varkappa_0({\bf
r}_1,{\bf r}_2) = \sum_{i,\lambda_1,\lambda_2}^{(0)}
\left(n_{\lambda_1}\,u_{i}^{\lambda_1}\,v_{i}^{\lambda_2}+
(1-n_{\lambda_1})\,u_{i}^{\lambda_2}\,v_{i}^{\lambda_1}\right)
\phi_{\lambda_1}({\bf r}_1) \phi_{\lambda_2}({\bf r}_2),
\label{kap01} \eeq where $n_{\lambda}=(0;1)$ are the occupation
numbers without pairing. The superscript ``(0)'' in the sum means
that all the states $\lambda$  belong to the model subspace:
$\eps_{\lambda} < E_0$. It can be easily seen that in the limit of
$E_0 \to \infty$, when the set \{$\lambda$\} is complete, the
expression (\ref{kap01}) coincides with Eq.~(\ref{kapuv}).

We consider a slab of nuclear matter embedded into the
one-dimensional potential  $U(x)$.  In this case, the momentum
${\bf k}_{\perp}$ in the $(y,z)$-plane (or ${\bf s}$-plane) is an
integral of motion. The single-particle wave functions can be
written as
\beq \phi_{\lambda}({\bf r}) = e^{i{\bf k}_{\perp}{\bf s}} y_n(x),
\label{fi} \eeq and the Bogolyubov functions have the form:
\beq u_i({\bf r}) = e^{i{\bf k}\pr{\bf s}} \,u_i(k\pr,x)\, \eeq
\beq v_i({\bf r}) = e^{i{\bf k}\pr{\bf s}} \,v_i(k\pr,x)\,,
\label{uvx} \eeq

Just as in Ref.~\cite{Pankr}, we will use the mixed
representation, i.e., the coordinate representation for the
$x$-direction and the momentum one for the  ${\bf s}$-plane. The
scheme of solving the gap equation in the mixed representation was
developed in detail in \cite{Pankr} for the case of the separable
form of the Paris potential \cite{Par1,Par2}.  Now we consider the
original form of the Argonne v$_{18}$ potential \cite{A18} and
present the main formulas for the general case.

Let us begin with the EPI. In the Wigner representation it depends
on the relative momenta $\bk_1$ and $\bk_2$  and on the CM
coordinates $X_1$ and $X_2$ in the input and the output channel.
As far as we deal with the  singlet pairing, the zeroth harmonic
of the EPI enters into the gap equation which depends only on the
absolute value of the relative momenta. One more advantage of the
two-step method is the possibility of using the so-called local
potential approximation (LPA) \cite{Rep} to find the EPI.
According to the LPA, the EPI at each point $X = (X_1 + X_2)/2$
can be replaced by the one in infinite nuclear matter placed in
the external potential field $U(X)$:
\beq \LPA(k_1,k_2;X_1,X_2)=\Vinf(k_1,k_2;t,[U(X)]),\label{LPA}
\eeq
where $t{=}X_1-X_2$. To find the EPI in infinite system, it is
convenient first to solve  the Lippman-Schwinger equation in the
complementary subspace in momentum space,
\beq \Vinf(k_1,k_2,P;E) = {\cal V}(k_1,k_2) + \int_{{\bf k}\in S'}
\frac{d^3k}{(2\pi)^3}\frac{{\cal V}(k_1,k)\Vinf(k,k_2,P;E)} {E
-\eps({\bf P},\bk)-\eps'({\bf P},\bk)},\label{inf} \eeq
where $E=2\mu$, $\eps({\bf P},\bk) = ({\bf P}/2 + \bk)^2/(2m) +
U(X)$, $\eps'({\bf P},\bk) = ({\bf P}/2 - \bk)^2/(2m) + U(X)$. The
integration in  (\ref{inf}) is carried out over $\bk\in S'$. The
quantity $\Vinf(k_1,k_2;t,[U(X)])$ is found from
$\Vinf(k_1,k_2;P,[U(X)])$ with the inverse Fourier transformation.

The explicit form of the gap equation (\ref{gap_kap}) in the
Wigner representation for the 1D geometry is as follows:
\beq \Delta(k,X) =
-\int\,\Vef(k,k';X,X')\;\varkappa_0(k'_{\pr},k'_{x};X')\;\frac{d^2\bk'_{\pr}}{(2\pi)^2}\,
\frac{dk'_{x}}{2\pi}\,dX'\,,\label{DeltawithXi}\eeq
The anomalous density can be written as
\beq\varkappa_0(k\pr,k_x;X) =
\sum^{(0)}_{nn'}\varkappa_0^{nn'}(k\pr)\;f_{nn'}(k_x,X)\,,\label{kred}\eeq
where \beq f_{nn'}(k_x,X) = \int
e^{-i\,k_x\,x}\,y_{n}(X+x/2)\,y_{n'}(X-x/2)\,dx\,,\label{f2}\eeq
and \beq \varkappa_{0}^{nn'}(k\pr) = \sum_{i}^{(0)}\left(n_{k\pr
n}\,u_{i}^{n}(k\pr)\,v_{i}^{n'}(k\pr)+ (1-n_{k\pr
n})\,u_{i}^{n'}(k\pr)\,v_{i}^{n}(k\pr)\right)\,.\label{kappall'}\eeq

Here we used the expansion of the Bogolyubov functions into the
basis \{$y_n$\}:
\beq u_i(k\pr,x) = \sum^{(0)}_{n}u^{n}_{i}(k\pr)\,y_{n}(x)\,,\quad
v_i(k\pr,x) =
\sum^{(0)}_{n}v^{n}_{i}(k\pr)\,y_{n}(x)\,.\label{uvred}\eeq

In this representation, the Bogolyubov equations have the form:
\beq\left\{\begin{aligned}
(\varepsilon_{nk\pr}-\mu)\;u_{i}^{n}(k\pr)+
\sum_{n'}\Delta_{nn'}(k\pr)\;v_{i}^{n'}(k\pr)
=E_{i}\,u_{i}^{n}(k\pr)\,,\\
\sum_{n'}\Delta_{nn'}(k\pr)\;u_{i}^{n'}(k\pr)
-(\varepsilon_{nk\pr}-\mu)\;v_{i}^{n}(k\pr)
=E_{i}\,v_{i}^{n}(k\pr)\,,\end{aligned}\right.\label{Beq2}\eeq
where $\varepsilon_{nk\pr}{=}\varepsilon_n+{k\pr}^2/2m$ and
\bea \Delta_{nn'}(k\pr) &=& \int
\Delta(k\pr,x,X)\;y_{n}(X+\frac{x}{2})\;y_{n'}(X-\frac{x}{2})\;dx\,dX=\notag\\
&=& \int
\Delta(k,X)\;f_{nn'}(k_x,X)\,dX\,\frac{dk_x}{2\pi}\,.\label{dred}\eea

The chemical potential $\mu$ in (\ref{Beq2}) is determined
inverting the equation for the number of particles per a unit
surface of the slab:
\beq \sigma = \int
\frac{d^{2}\bk\pr}{(2\pi)^{2}}\,\sum_{i\lambda}\,
\left(v_{i}^{\lambda}(k\pr)\right)^{2} =
\mathrm{const}\,.\label{sigma}\eeq

The above relations determine the gap function in the slab system.

\section{Calculation results}

Just as in \cite{Pankr}, we use the Saxon-Woods potential $U(x)$
symmetrical with respect to the point $x=0$ with typical for
finite nuclei values of the potential well depth $U_0=-50\;$MeV
and diffuseness parameter of $d=0.65\;$fm. For the main part of
calculations the thickness parameter of the slab was chosen as
$L=6\;$fm to mimic nuclei of the tin region. As far as the
Hamiltonian of the system is symmetrical under the axis reflection
$x \to -x$, the eigenfunctions $y_n$ can be separated into even,
$y_n^+$, and odd, $y_n^-$, functions. It simplifies the numerical
procedure essentially, as far the $n,n'$ indices  in (\ref{Beq2})
should numerate the states of a fixed parity.

The calculation scheme is rather different from that used in
\cite{Pankr} for the separable form of the $NN$-force. The common
feature is the convenience to use the scheme of parallel
calculations for different $k\pr$ momentum, since the slab is
homogeneous in the ${\bf s}-$plane. The total number of processors
we used is 210, which gives required accuracy for calculations
with the model space $S_0$ specified by $E_0=15\;$MeV. The
discrete spectrum representation method was used to take into
account the continuum, $0<\eps_\lambda<E_0$. The infinite wall was
put at the distance $R=25\;$fm. The independence of results on the
value of $R$ was checked. The gap equation $(\ref{DeltawithXi})$
was solved with iterations. The procedure was stopped when the
maximal difference of values of $\Delta(k,X)$ obtained in two
subsequent iterations became less than 10$^{-6}\;$MeV. The typical
number of effective iterations was about 10. The chemical
potential was found anew according (\ref{sigma}) on each
iteration. The self-consistent value of the chemical potential
$\mu{=}-7.95\;$MeV is very close to the one, $\mu_0{=}-8\;$MeV,
without pairing. The difference $\delta \mu{=}-0.05\;$MeV is in
agreement with the standard estimate $\delta \mu \sim -
\Delta^2/\eps_{\Fs}$.

To illustrate the solution graphically, we calculated the ``Fermi
averaged'' gap
\beq
\DF(X)= \Delta(k_{\Fs}(X),X). \label{Del_F}
\eeq
Here the local Fermi momentum is defined as follows:
$k^2_{\Fs}(X){=}2m(\mu{-}U(X))$ if  $2m(\mu{-}U(X)) > 0$, and
$k^2_{\Fs}(X){=}0$ otherwise. Analogously, we could define the
Fermi averaged EPI as
\beq \VF (X) = \int dt \Vef (k_1{=}k_2{=}k_{\Fs}(X);t,X).
 \label{VF} \eeq

\begin{figure}[h]\vspace{-2mm}
\begin{center}
\includegraphics[height=80mm,width=100mm]{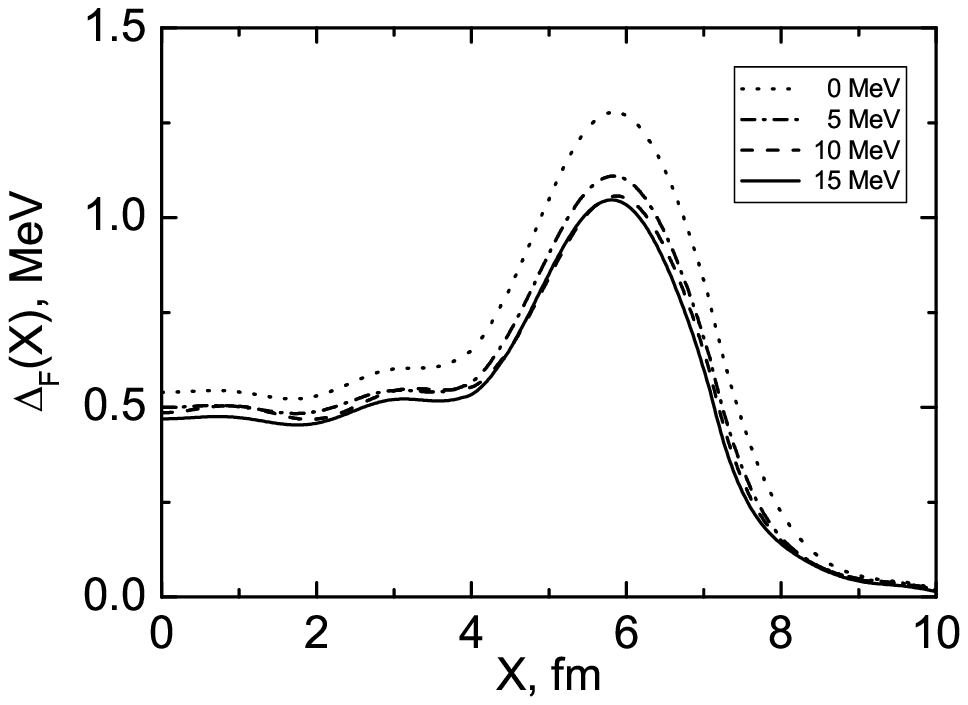}%
\vspace{0mm}
\end{center}
\caption{ The Fermi averaged gap $\DF(X)$ for different $E_{0}$.}
\label{Del_E0}
\end{figure}

There are two parameters which influence essentially difficulties
of the calculation procedure. They are the separation energy $E_0$
which defines the model space $S_0$ and the cut-off momentum
$K_{\rm max}$ in the integral of Eq.~(\ref{inf}). Let us begin
from choosing the first one. Fig.~\ref{Del_E0} shows the
dependence of the solution on the value of $E_0$. Analyzing these
curves one can see that the variation of $E_0$ from 5 MeV to 10
MeV results in the change of the maximum of the Fermi-averaged gap
function of the order of 5\% whereas the next step to 15 MeV
diminishes this difference to 1\%. A more quantitative information
could be obtained from the matrix elements of the gap. The
diagonal matrix elements $\Delta_{nn}(k\pr{=}0)$ for different
$E_0$ are presented in Table 1. The even  states are located in
the upper part of the table, the odd ones, in the lower part. One
can see that the convergence of $\Delta_{nn}$ values increasing
$E_0$ is slower than that of the maximum of $\DF(X)$. Now the
average change of a matrix element with the variation of $E_0$
from 10 MeV to 15 MeV is about 3\%. This is, of course, the upper
estimate of the accuracy if we choose $E_0=15\;$MeV as an
appropriate value. Analyzing the convergence of  $\Delta_{nn}$
values in Table 1, one can suppose that the change under
consideration will be about 1-2\% at the next step to
$E_0=20\;$MeV. Therefore more realistic estimate of the accuracy
of the calculations for $E_0=15\;$MeV is 1-2\%.

\begin{table}[t!]
\caption{ Diagonal matrix elements
$\Delta_{nn}(k_{\pr}{=}0)\;$(MeV) in the slab with the width
parameter $L{=}6\;$fm for Argonne v$_{18}$ potential for different
$E_0\;$(MeV). $\eps_n\;$(MeV) is the single particle spectrum.}
\begin{center}

\begin{tabular}{|c|c|}
\hline  \hspace*{3.2ex}  $\eps_n$ \hspace*{3.2ex} & \hspace*{27ex} $\Delta_{nn}$ \hspace*{27ex} \\
\end{tabular}
\vskip 0 mm

\begin{tabular}{|c|c|c|c|c|}
\hline  \hspace*{10ex} & \hspace*{2ex} $E_0$ = 0 \hspace*{2ex} &
\hspace*{2ex} $E_0$ = 5 \hspace*{2ex}
& \hspace*{2ex} $E_0$ = 10 \hspace*{2ex} & \hspace*{2ex} $E_0$ = 15 \hspace*{2ex} \\

\hline
-48.26 & 1.516 & 1.362 & 1.339 & 1.284 \\
-36.55 & 1.461 & 1.282 & 1.245 & 1.210 \\
-18.60 & 1.198 & 1.034 & 0.982 & 0.972 \\
-1.325 & 0.421 & 0.342 & 0.324 & 0.309 \\
\hline
\hline
-43.51 & 1.522 & 1.339 & 1.309 & 1.266 \\
-28.04 & 1.353 & 1.174 & 1.119 & 1.102 \\
 -9.10 & 0.903 & 0.764 & 0.747 & 0.717 \\
\hline

\end{tabular}\label{tab_1}
\end{center}
\end{table}

The latter could be referred to both approximations used in the
version of the the two-step approach outlined in Sect.~2. These
are: neglecting of the pairing in the complementary space $S'$ and
using the LPA as well. Note that in the case of the Paris
potential 1\% accuracy was reached at $E_0=20\;$MeV \cite{Pankr}.

\begin{figure}[h!]\vspace{-2mm}
\begin{center}
\includegraphics[height=70mm,width=140mm]{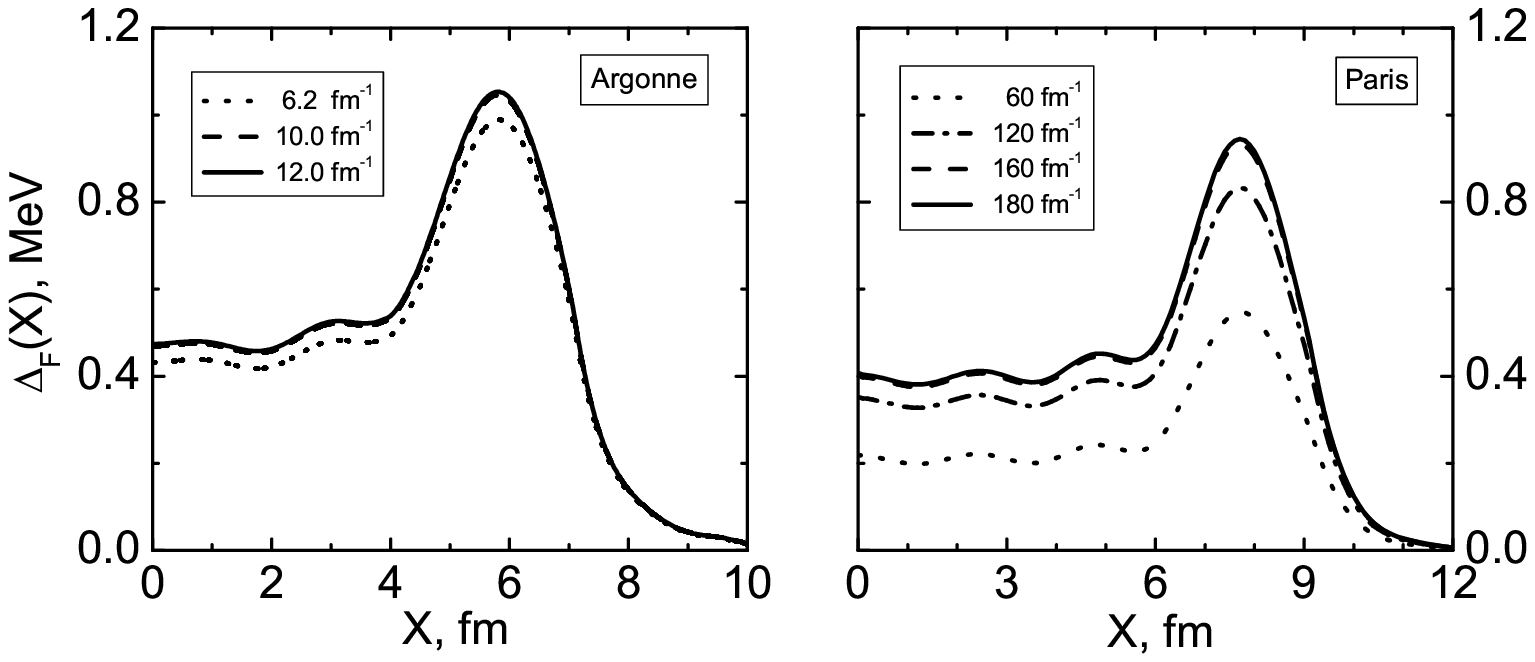}%
\vspace{0mm}
\end{center}
\hskip 1.5 cm \caption{ The Fermi averaged gap $\DF(X)$ for
different cut-off momenta $K_{\rm max}$ (in fm$^{-1}$) for the
Argonne $v_{18}$ potential (the left panel, $E_0=15$ MeV) and the
Paris potential (the right panel, $E_0=20$ MeV).} \label{Del_km}
\end{figure}

Let us turn now to the analysis of dependence of the gap function
on the cut-off momenta $K_{\rm max}$ in Eq.~(\ref{inf}). The
Fermi-averaged gap $\DF(X)$ is displayed in the left panel of
Fig.~\ref{Del_km}  for different $K_{\rm max}$ in the case of the
Argonne $v_{18}$ potential. The value of $K_{\rm
max}=6.2\;$fm$^{-1}$ corresponds to the cut-off energy $E_{\rm
max}=800\;$MeV used in \cite{milan2}. We see that this value of
$K_{\rm max}$ guarantees only 10\% accuracy of the gap value and
only $K_{\rm max}=10\;$fm$^{-1}$ is sufficiently big to result in
1\% accuracy. Therefore it is possible that the gap found in
\cite{milan2} is underestimated at approximately 10\% due to not
sufficiently big value of $E_{\rm max}$. Note that another version
of the Argonne force, $v_{14}$, was used there but it is quite
close to that used by us. In the systematic calculations, the
values of   $E_0=15\;$MeV and $K_{\rm max}=10\;$fm$^{-1}$ will be
chosen.

For comparison, the analogous dependence is shown for the Paris
potential in the right panel of Fig.~\ref{Del_km} demonstrating
very slow convergence of the integral in Eq.~(\ref{inf}) in this
case. Indeed, now the huge value of $K_{\rm max}=160\;$fm$^{-1}$
is necessary to guarantee 1\% accuracy for the Paris potential.
Only the use of the separable version of the Paris force in
\cite{Pankr} made it possible to carry out calculations with
sufficient accuracy.

\begin{figure}[h!]\vspace{-2mm}
\begin{center}
\includegraphics[height=70mm,width=140mm]{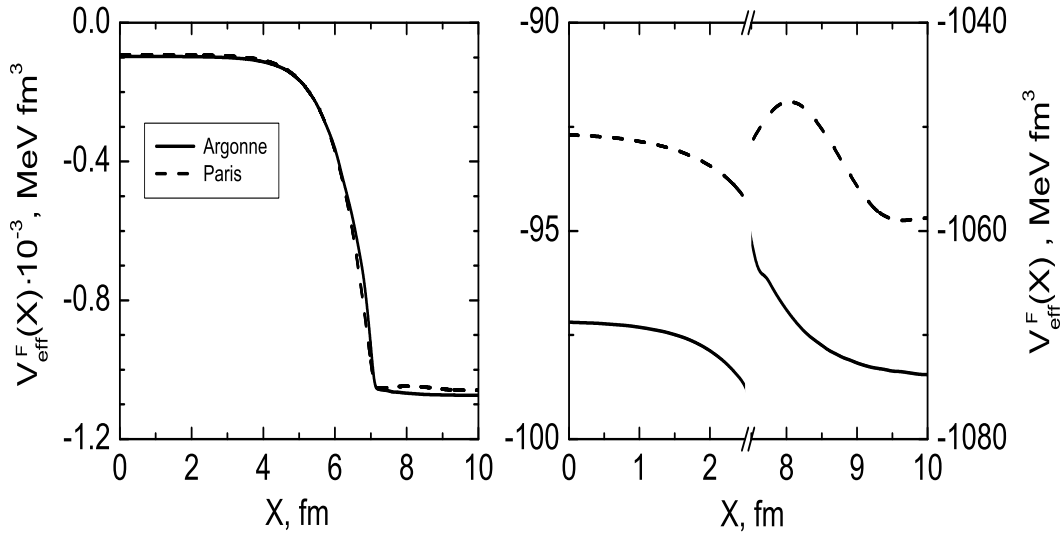}%
\vspace{0mm}
\end{center}
\hskip 1.5 cm
\begin{minipage}[c]{0.8\linewidth}
\caption{ The Fermi averaged EPI  $\VF(X)$ for the Argonne
$v_{18}$ potential (solid line) and the Paris potential (dashes),
 $E_0=20$ MeV) for both the cases.}
\end{minipage}
\label{EPI_Par_Arg}
\end{figure}

\begin{figure}[h!]\vspace{-2mm}
\begin{center}
\includegraphics[height=80mm,width=100mm]{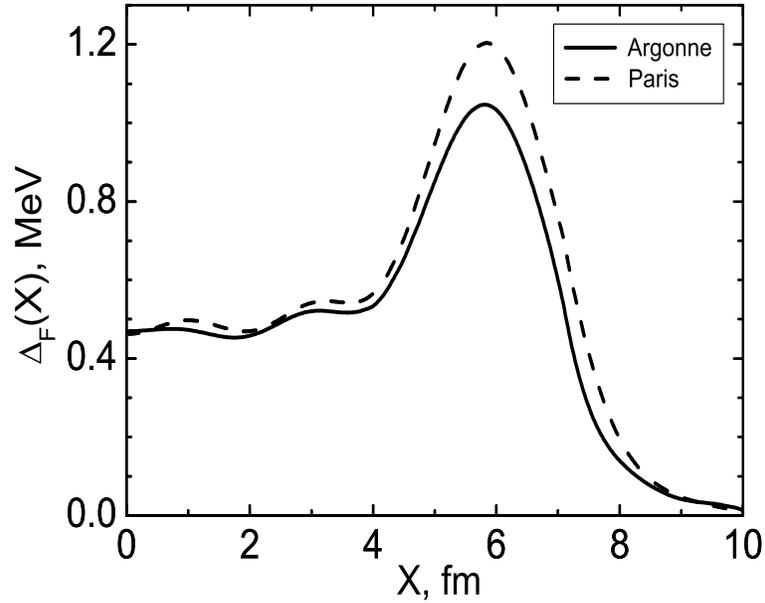}%
\vspace{0mm}
\end{center}
\caption{ The Fermi averaged gap $\DF(X)$ for the Argonne $v_{18}$
potential (solid line) and the Paris potential (dashes).}
\label{Del_fin}
\end{figure}

Fig.~\ref{EPI_Par_Arg} demonstrates that the EPI found from
Eq.~(\ref{inf}) depends only a little on the specific form of the
realistic $NN$-potential used provided the convergence of the
integral in the momentum space is reached. One can see that the
two curves for different potentials are nearly coincident. To show
a small difference we displayed the same curves at small and large
values of $x$ separately in a large scale. We see that the
difference is about 2\% in the asymptotic region where the EPI is
big and about 5\% where the EPI is small. However, even such a
small variation of the EPI is not negligible as far as any change
of the pairing force is significantly enhanced in the gap
equation. As the analysis of \cite{Pankr} has shown, in the system
under consideration a 1\% variation of the EPI results in
approximately 5\% variation of the gap. It is confirmed with
direct comparison of the gap functions for both the potentials
which is presented in Fig.~\ref{Del_fin}. The maximual difference
between values of $\DF(X)$ for different potentials is of the
order of 10\%.

To analyze this difference quantitatively it is instructive to
compare the  matrix elements of the gap. They are presented in
Table 2 for the  diagonal matrix elements $\Delta_{nn}(k\pr{=}0)$.
One can see that the difference under consideration is again of
the order of 10\%. The only exception is the state with very small
energy $\eps_n{=}-1.3\;$MeV for which the matrix element of
$\Delta$ itself is essentially less than the typical values which
is of the order of 1 MeV. This peculiarity of states with small
energy is caused with a special form of their wave functions which
have very long ``tails'' outside the slab. Therefore the weight of
$y_n(x)^2$ in the integral of $\Delta_{nn}$ is rather small in the
region where the $\Delta(X)$ is large. The upper half of the table
contains the positive parity states. These are the matrix elements
which could be related to those in heavy nuclei over the
single-particle $s$-states. Indeed, the $k\pr$ variable in a slab
is an analogue of the orbital angular momentum $l$ in a spherical
nucleus, $k\pr={0}$ corresponding to $l={0}$. The average value of
$\Delta_{nn}(k\pr{=}0)$ for the positive parity states is
$\bar{\Delta}{=}0.94\;$MeV for the Argonne force and
${=}1.04\;$MeV  for the Paris potential. It looks reasonable to
exclude the state with anomalously small energy for calculating
the average value of the gap. In this case, one finds
$\bar{\Delta}{=}1.16\;$MeV for the Argonne force and
${=}1.27\;$MeV for the Paris potential. These values should be
compared with the experimental values of the gap in the tin region
which are of the order of $\Delta_{\rm exp}\simeq 1.3\;$MeV. We
see that although the gap value for the Argonne force is a little
(10\%) less than that for the Paris force it is also rather close
to the experimental data, reserving only narrow room ($\simeq
10-20$\%) for contributions of the surface vibrations.

\begin{table}[t]
\caption{ Diagonal matrix elements
$\Delta_{nn}(k_{\pr}{=}0)\;$(MeV) in the slab with the width
parameter $L{=}6\;$fm. $\eps_n\;$(MeV) is the single particle
spectrum.}
\begin{center}

\begin{tabular}{|c|c|}
\hline  \hspace*{3.5ex}  $\eps_n$ \hspace*{3.1ex} & \hspace*{27.2ex} $\Delta_{nn}$ \hspace*{27ex} \\
\end{tabular}
\vskip 0 mm

\begin{tabular}{|c|c|c|}
\hline \hspace*{10.3ex} &  \hspace*{4.5ex} Argonne v$_{18}$
potential \hspace*{4.5ex} & \hspace*{4.5ex} Paris potential \hspace*{4.5ex} \\

\hline
-48.26 & 1.284 & 1.396 \\
-36.55 & 1.210 & 1.326 \\
-18.60 & 0.972 & 1.077 \\
-1.325 & 0.309 & 0.357 \\
\hline
\hline
-43.51 &  1.266 & 1.385 \\
-28.04 &  1.119 & 1.227 \\
 -9.10 &  0.764 & 0.817 \\
\hline

\end{tabular}\label{tab_2}
\end{center}
\end{table}

The quantity $\DF(X)$ is a localized representation of the gap. To
examine the nonlocal properties of the gap, it is reasonable to
calculate the $X$-dependent rms radius
\beq r_{\Delta}(X) = \sqrt{\langle r^2_{\Delta}(X)\rangle}\,,\eeq
where
\beq \langle r^2_{\Delta}(X)\rangle = \dfrac{\int{\bf
r}^2\Delta(r,X)\;d^3\br}{\int\Delta(r,X)\;d^3\br} =
\left.\dfrac{-\dfrac{\partial^2}{\partial
k^2}\Delta(k,X)}{\Delta(k,X)}\right|_{k=0}.\eeq

\begin{figure}[h!]\vspace{-2mm}
\begin{center}
\includegraphics[height=80mm,width=100mm]{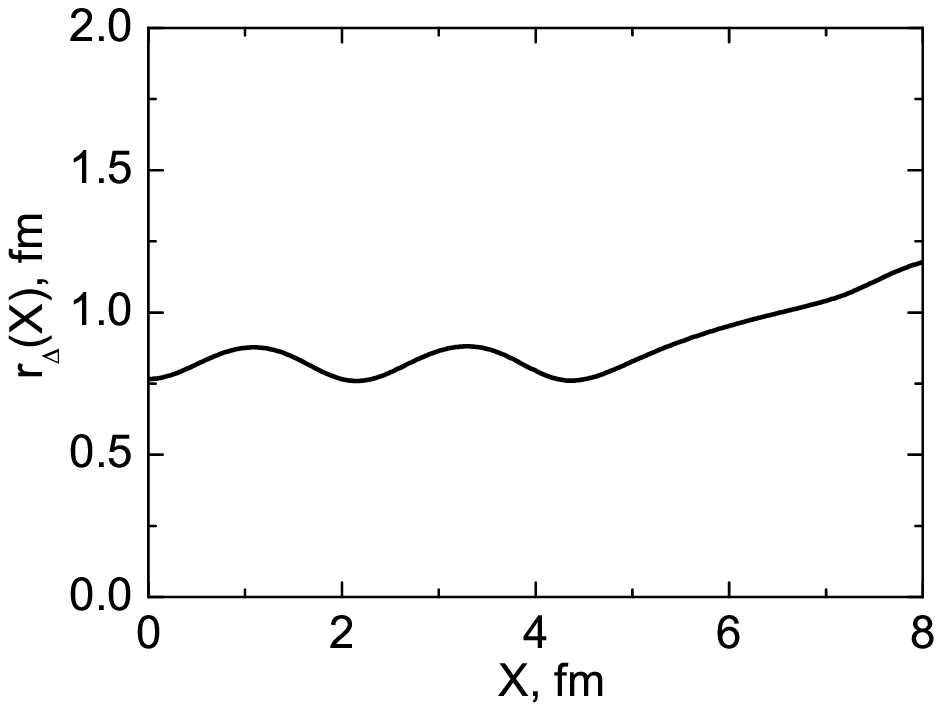}%
\vspace{0mm}
\end{center}
\caption{ The rms radius of the gap $\DF(X)$ for the Argonne
$v_{18}$ potential.} \label{rsq_del}
\end{figure}

It is displayed in Fig.~\ref{rsq_del}. We see that it is about 0.8
fm inside the slab and grows to $\simeq 1.2\;$fm outside, i.e. the
gap function is essentially non-local.

\section{On the $\mu$-dependence of the gap}

In this section, we analyze dependence of the pairing gap on the
chemical potential $\mu$. This subject is of interest due to a
progress in physics of of nuclei laying far off the
$\beta$-stability valley reached in recent years.  In particular,
it concerns  nuclei in the vicinity of the nucleon drip line,
where the chemical potential $\mu$  of the neutron or the proton
subsystem vanishes. The importance of nucleon pairing for this
region is generally accepted.  In studies aimed at calculating the
nucleon drip line within the Hartree--Fock method \cite{Dob1,Dob2}
or the generalized EDF method \cite{Fay}, which employ a
phenomenological nucleon--nucleon interaction, much attention was
given to exploring special features of pairing at small values of
$\mu$. In this connection, the main interest was put to such
problems as the correct taking into account continuum states, as a
comparison of the exact solution of the Bogolyubov equations with
that found within the BCS approximation, and so on. However, the
possibility that the parameters of the EPI determining $\Delta$
change in the vicinity of the drip line has been ignored so far.
It is hardly possible to address this issue within a
phenomenological approach. This requires the use of the {\it ab
initio} methods which start from a free NN-interaction.

Such a study was carried out in \cite{mu} for the Paris potential.
Dependence of $\Delta$ on $\mu$ turned out to be model dependent
provided  the mean field $U(x)$ is not found self-consistently.
The matter is that, as it was shown in \cite{BLSZ1,ST}, the
potential well $U(x)$ could change in vicinity of the drip point
significantly due to dependence of the scalar Landau--Migdal
amplitudes \cite{AB} on $\mu$. As far as this problem is not
solved up to now, two models were examined in \cite{mu}. In the
first one (Model 1), the depth of the potential well
$U_0{=}{-}50\;$MeV is fixed, the central density increasing with
decrease of $|\mu|$. In the second model (Model 2), the central
density is considered to be a constant, the value of $|U_0|$
decreasing with decrease of $|\mu|$.  It turned out that, with
decrease of $|\mu|$, the gap value decreases in the first model
and increases in the second one. In addition, the $\mu$ dependence
of the so-called gap-shape function was examined in \cite{mu}.
This quantity is defined as the ratio \beq\chi_{\mbox{\scriptsize
F}}(X)=\frac {\DF(X)} {\DF(0)}. \label{xi}\eeq It determines
directly the effect of the surface enhancement of the gap
\cite{BLSZ2,6auth}. It turned out that this effect grows with
decrease of $|\mu|$ within both the models.

\begin{figure}[t!]\vspace{-2mm}
\begin{center}
\includegraphics[height=70mm,width=140mm]{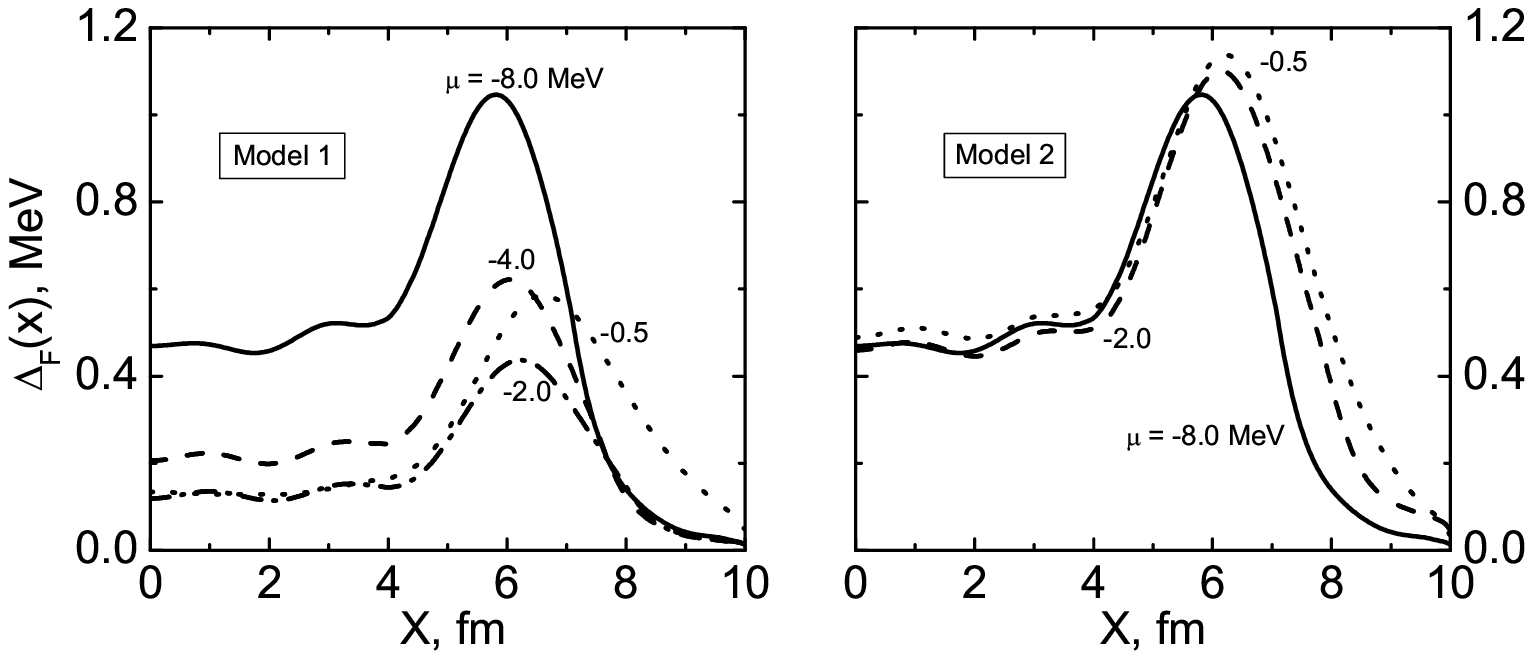}%
\vspace{0mm}
\end{center}
\caption{ The Fermi averaged gap $\DF(X)$ for different values of
the chemical potential $\mu$.} \label{Del_mu}
\end{figure}

\begin{figure}[t!]\vspace{-2mm}
\begin{center}
\includegraphics[scale=0.85,keepaspectratio=true]{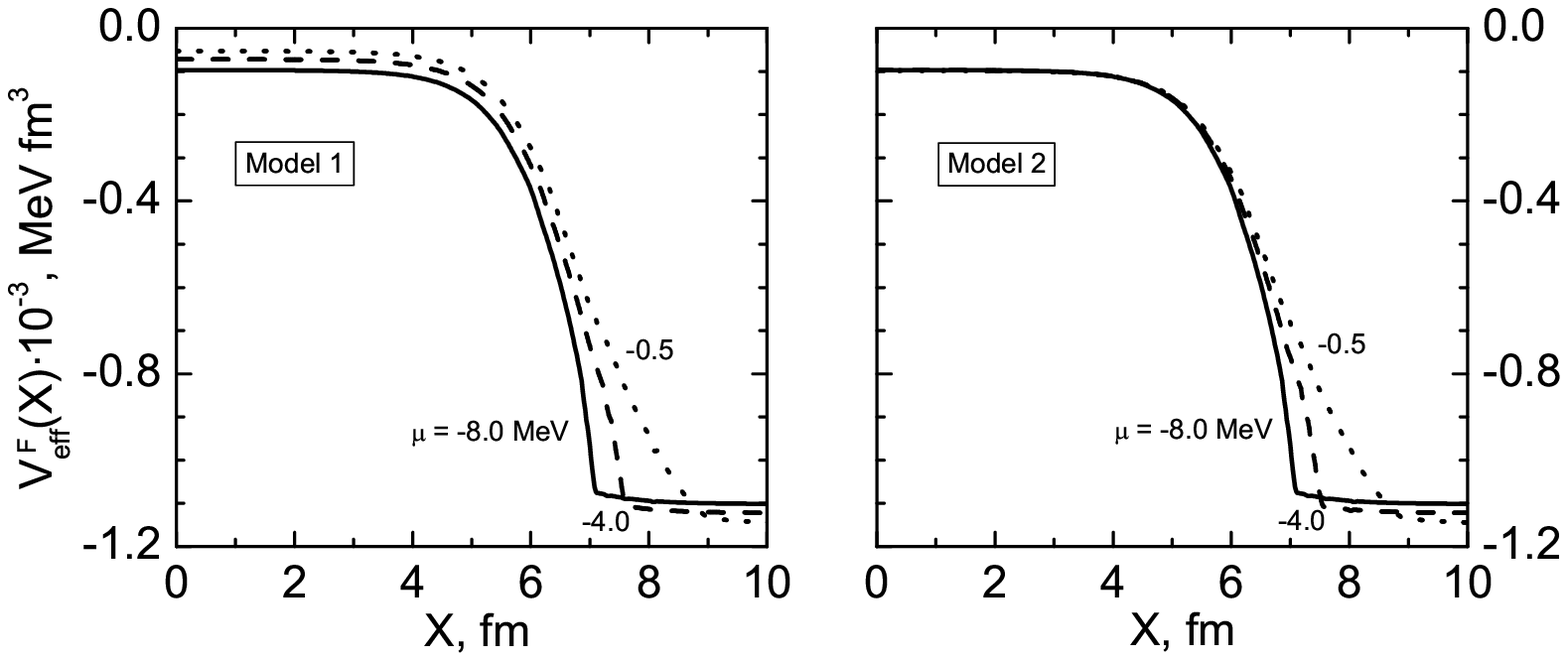}%
\vspace{0mm}
\end{center}
\vspace{-5mm} \caption{ The Fermi averaged EPI $\VF(X)$ for
different values of the chemical potential $\mu$.} \label{Vef_mu}
\end{figure}

Now we carry out analogous calculations for the Argonne force
v$_{18}$. Dependence of the gap on the chemical potential for both
the models is displayed in Fig.~\ref{Del_mu}. In general, it
repeats the behavior of the gap in the case of the Paris potential
\cite{mu}. The main origin of the effect is the $\mu$-dependence
of the EPI which is shown in Fig.~\ref{Vef_mu}. In the first
model, with decrease of $|\mu|$, the EPI becomes a little stronger
in the asymptotic region and weaker inside the slab. The first
effect originates from closeness to the pole in the free
$NN$-scattering amplitude, the second one, due to growing of the
local value of the Fermi momentum. It turned out that, with
diminishing of $|\mu|$, the second reason dominates and $\Delta$
falls, and only at very small $|\mu|$ the surface enhancement of
the EPI overcomes. In the second model, the EPI inside the slab
doesn't practically depend on $\mu$, and only the surface effect
remains. Therefore in this case the gap value is growing with
$|\mu|$ decreasing, but the effect is rather small.

\begin{figure}[t!]\vspace{-2mm}
\includegraphics[height=60mm,width=135mm]{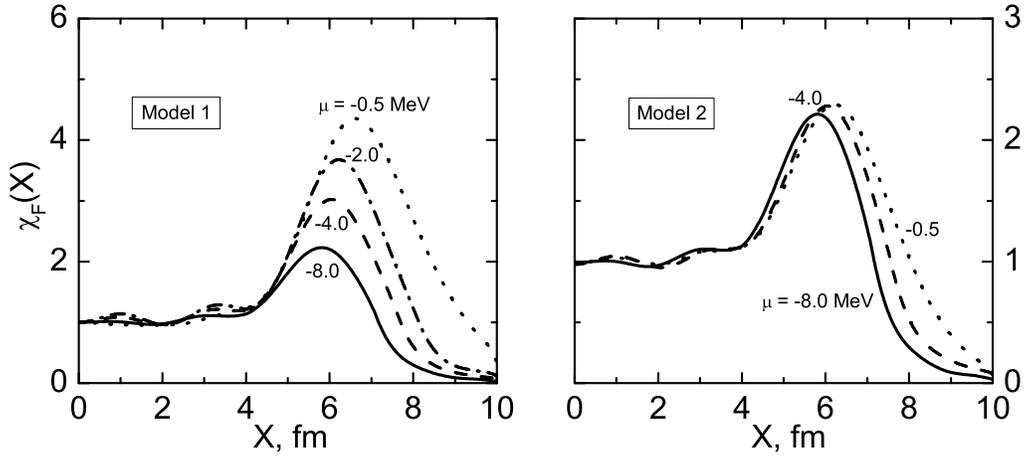}%
\vspace{0mm}
\caption{ The gap shape function $\chi(X)$ for different values of
the chemical potential $\mu$.} \label{Xi_mu}
\end{figure}

In the first model, the $\mu$-dependence effect is much more
pronounced for the gap-shape function, see Fig.~\ref{Xi_mu}, left
panel. In this case, the surface enhancement of the gap becomes
stronger with decrease of $|\mu|$, the effect being monotonic. It
occurs because the nominator in (\ref{xi}) diminishes slower than
the denominator. In the second model, the denominator in
(\ref{xi}) almost doesn't depend on $\mu$, and qualitatively the
$\mu$-dependence of gap-shape function repeats that of the gap
itself.

\section{Conclusions}

We solved the microscopic gap equation for $^1S_0$ pairing in a
nuclear slab for the Argonne v$_{18}$ $NN$-potential. The slab is
embedded into the Saxon-Woods potential well $U(x)$ with
parameters typical for heavy nuclei. The width parameter $L=6\;$fm
was chosen to mimic nuclei of the tin region examined in
\cite{milan2}. The features of the gap function obtained are very
close to those found previously for the separable form of the
Paris potential \cite{Pankr}, but the absolute value of $\Delta$
turned out to be 10\% less. This difference should be attributed
to essentially different form of the free $NN$-potentials used
which is the main input of the microscopic gap equation.
Evidently, it may be considered as the upper estimate of the
uncertainty originating from the specific choice of the
$NN$-potential. Indeed, the two potentials are very contrast to
each other, the Argonne force being rather soft and the Paris one,
very hard. The diagonal matrix elements of $\Delta$ obtained are
about 1.1 MeV which could be compared with the experimental value
of $\Delta\simeq 1.3\;$MeV in tin isotopes with $A\simeq 120$. On
the other hand, the gap value obtained is significantly bigger
than $\Delta\simeq 0.7\;$MeV found in \cite{milan2} where the lack
of the gap was attributed to the contribution of the surface
vibrations. Omitting the discussion of possible uncertainties in
calculating these corrections (see, e.g., \cite{Rep}), we mention
only that  the contribution of about 20\% was obtained in
\cite{Kam} for the process under consideration. This estimate
agrees with our calculations.

We think that different values of the effective nucleon mass $m^*$
used is the main reason of the contradiction  under discussion. In
\cite{milan2} the coordinate dependent effective mass $m^*(r)$ was
used from the Sly4 force \cite{Sly4} which is essentially less
than the bare nucleon mass, $m^*=m$, used by us. The latter is
chosen in accordance with prescriptions of \cite{KhS} and
\cite{Fay} which are based, in particular, on the analysis of the
single-particle spectra of magic nuclei. Indeed, it is well known
that the use of $m^* < m$ in any Skyrme-type forces results in the
single-particle spectra which are significantly decompressed.

We analyzed also the dependence of the gap on the chemical
potential $\mu$. The result turned out to be model dependent in
the approach with the potential well $U(x)$ considered as an input
of the calculation scheme. In the first model, the depth
$U_0{=}{-}50\;$MeV is fixed, the central density increasing with
decrease of $|\mu|$. In this case, the value of $\Delta$ decreases
with diminishing of $|\mu|$ and only at very vicinity of the drip
point it begins to grow. A different behavior was found of the
gap-shape function which characterizes the surface enhancement of
the gap. This quantity turned out to be monotonically increasing
with decrease of $|\mu|$. In the second model, the central density
is supposed to be  constant, the depth of $U(x)$ decreasing with
decrease of $|\mu|$.  In this case, the $\mu$-dependence effect is
less pronounced but the gap value and gap-shape function grow both
with decrease of $|\mu|$. Thus, a self-consistent calculation of
the potential well depending on $\mu$ is necessary to clarify the
properties of nuclear pairing in the drip line vicinity. In this
case, a $\mu$-dependence of the effective force should be taken
into account \cite{BLSZ1,ST} if one deals with the
phenomenological approach.

\section{Acknowlegements}

The authors are highly thankful to S.T.~Belyaev and
S.V.~Tolokonnikov for helpful discussions. This research was
partially supported by the Grant NSh-8756.2006.2 of the Russian
Ministry for Science and Education and by the RFBR grants
06-02-17171-a and 07-02-00553-a.


\newpage
\begin{thebibliography}{0}

\bibitem{milan1}
 F.~Barranco, R.~A.~Broglia, H.~Esbensen, and E.~Vigezzi,
 Phys. Lett. B {\bf 390} (1997) 13.

\bibitem{milan2} F.~Barranco, R.~A.~Broglia, G.~Colo, G.~Gori,
E.~Vigezzi, and P.~F.~Bortignon. Eur.~Phys.~J.~A 21 (2004) 57.

\bibitem{Pankr}
 S.~S.~Pankratov, M.~Baldo, U.~Lombardo, E.~E.~Saperstein, and M.~V.~Zverev,
  Nucl. Phys. A 765 (2006) 61.

\bibitem{Paris} M.~Lacombe, B.~Loiseaux, J.~M.~Richard, R.~Vinh Mau,
J.~C\^ot\'e, D.~Pir\`es and R.~de Tourreil, Phys.~Rev.~C\,21
(1980) 861.

\bibitem{Bal} M.~Baldo, J.~Cugnon, A.~Lejeune, U.~Lombardo,
   Nucl.~Phys.~A\,515 (1990) 409.

\bibitem{Sly4} E.~Chabanat, P.~Bonche, P.~Haensel, J.~Meyer,
and R.~Schaeffer,  Nucl.~Phys.~A 627 (1997) 710.

\bibitem{KhS} V.~A.~Khodel and E.~E.~Saperstein, Phys.~Rep. 92 (1982) 183.

\bibitem{Fay}
S.~A.~Fayans, S.~V.~Tolokonnikov, E.~L.~Trykov, and D.~Zawischa,
Nucl.~Phys.  A 676 (2000) 49.

\bibitem{A18}
 R.~B.~Wiringa, V.~G.~J.~Stoks, and R.~Schiavilla,   Phys. Rev.
C 51 (1995) 38.


\bibitem{Rep} M.~Baldo, U.~Lombardo, E.~E.~Saperstein, M.~V.~Zverev,
 Phys.~Rep. 391 261 (2004).

\bibitem{AB} A.~B.~Migdal, Theory of finite Fermi systems and applications to
atomic nuclei (Wiley, New York, 1967).


\bibitem{Schuck} P.~Ring, P.~Schuck, The nuclear many-body problem
(Springer, Berlin, 1980).


\bibitem{Par1} J.~Haidenbauer, W.~Plessas, Phys.~Rev.~C 30 (1984) 1822.

\bibitem{Par2} J.~Haidenbauer, W.~Plessas, Phys.~Rev.~C 32 (1985) 1424.

\bibitem{Dob1}
J.~Dobaczewski, H.~Flocard, and J.~Treiner, Nucl. Phys. A 422
(1984) 103.

\bibitem{Dob2}
J.~Dobaczewski {\it et al.}, Phys. Rev. C 53 (1996) 2809.

\bibitem{mu}
 S.~S.~Pankratov, E.~E.~Saperstein, and M.~V.~Zverev, Phys. At.
 Nucl. 69 (2006) 2009.

\bibitem{BLSZ1} M.~Baldo, U.~Lombardo, E.~E.~Saperstein, M.~V.~Zverev,
Phys. Lett. B 533 (2002) 17.

\bibitem{ST}  E.~E.~Saperstein, S.~V.~Tollokonnikov, JETP Lett,
 78 (2003) 795.


\bibitem{BLSZ2} M.~Baldo, U.~Lombardo, E.~E.~Saperstein, M.~V.~Zverev,
Phys. Lett. B 459 (1999) 437.

\bibitem{6auth} M.~Baldo, M.~Farine, U.~Lombardo, E.~E.~Saperstein,
P.~Schuck and M.~V.~Zverev,
 Eur.~Phys.~J.~A 18 (2003) 17.


\bibitem{Kam} A.~V.~Avdeenkov, S.~P.~Kamerdzhiev, JETP~Lett. 69 (1999)
715.


\end{thebibliography}
\end{document}